\documentclass[11pt]{article}

\usepackage{latexsym}
\usepackage{amsfonts}
\usepackage{latexsym}
\usepackage{graphicx}
\usepackage{amsmath}
\usepackage{amssymb}

\title{Bicomplex quantum mechanics:\\ I. The generalized Schr\"odinger equation}
\author{D. Rochon\thanks{E-mail: \texttt{dominic.rochon@uqtr.ca}} \and
S. Tremblay\thanks{E-mail: \texttt{sebastien.tremblay@uqtr.ca}}}
\date{D\'epartement de math\'ematiques et
d'informatique \\ Universit\'e du Qu\'ebec \`a Trois-Rivi\`eres \\
C.P. 500 Trois-Rivi\`eres, Qu\'ebec \\ Canada, G9A 5H7}

\makeatletter \@addtoreset{equation}{section}

\makeatletter

\def\al{\alpha}
\def\b{\beta}
\def\ga{\gamma}
\def\de{\delta}

\def\be   {\begin{equation}}   \def\ee   {\end{equation}}
\def\ba   {\begin{array}}      \def\ea   {\end{array}}
\def\bea  {\begin{eqnarray}}   \def\eea  {\end{eqnarray}}
\def\bean {\begin{eqnarray*}}  \def\eean {\end{eqnarray*}}

\newtheorem{theorem} {Theorem}

\newtheorem{corollary} {Corollary}

\newcommand{\me}{\mathrm{e}}

\newcommand{\pre}{\mathrm{Re}}
\newcommand{\pim}{\mathrm{Im}}

\newcommand{\bo} {\ensuremath{{\bf i_1}}}
\newcommand{\bos}{\ensuremath{{\bf i_1^{\text 2}}}}
\newcommand{\bt} {\ensuremath{{\bf i_2}}}
\newcommand{\bts}{\ensuremath{{\bf i_2^{\text 2}}}}
\newcommand{\bz} {\ensuremath{{\bf i_0}}}
\newcommand {\bj}{\ensuremath{{\bf j}}}
\newcommand {\bjs}{\ensuremath{{\bf j^{\text 2}}}}

\begin{document}
\maketitle
\begin{abstract}
We introduce the set of bicomplex numbers $\mathbb{T}$ which is a
commutative ring with zero divisors defined by
$\mathbb{T}=\{w_0+w_1 \bold{i_1}+w_2\bold{i_2}+w_3 \bold{j}|\
w_0,w_1,w_2,w_3 \in \mathbb{R}\}$ where $\bold{i^{\text 2}_1}=-1,\
\bold{i^{\text 2}_2}=-1,\ \bold{j}^2=1,\
\bold{i_1}\bold{i_2}=\bold{j}=\bold{i_2}\bold{i_1}$. We present
the conjugates and the moduli associated with the bicomplex
numbers. Then we study the bicomplex Schr\"odinger equation and
found the continuity equations. The discrete symmetries of the
system of equations describing the bicomplex Schr\"odinger
equation are obtained. Finally, we study the bicomplex Born
formulas under the discrete symetries. We obtain the standard
Born's formula for the class of bicomplex wave functions having a
null hyperbolic angle.

\vspace{1cm}\center{PACS: 03.65.-w, 02.10.Hh}
\end{abstract}

\newpage
\section{Introduction}
In this paper we investigate the bicomplex Schr\"odinger equation
where bicomplex numbers $\mathbb{T}$ (also called ``tetranumbers''
 in the literature) are defined as the set
$\mathbb{T}:=\{w_0\bold{i_0}+w_1 \bold{i_1}+w_2\bold{i_2}+w_3
\bold{j}|\ w_0,w_1,w_2,w_3 \in \mathbb{R}\}$ with
\begin{center}
\be
\begin{tabular}{|c||c|c|c|c|}
\hline
$\cdot$& \bz & \bo  & \bt  & \bj  \\
\hline
\hline
\bz    & \bz & \bo  & \bt  & \bj  \\
\hline
\bo    & \bo & -\bz & \bj  & -\bt \\
\hline
\bt    & \bt & \bj  & -\bz & -\bo \\
\hline
\bj    & \bj & -\bt & -\bo & \bz \\
\hline
\end{tabular}
\label{eq:ij}
\ee
\end{center}

We call $\bold{i_1}$ and $\bold{i_2}$ the imaginary units and we
attribute to $\bold{j}$ the name of hyperbolic (imaginary) unit.
The set of bicomplex numbers is a {\em commutative} ring with unit
and zero divisors. Hence, contrary to quaternions, bicomplex
numbers are commutative with some non-invertible elements situated
on the ``null cone''.

The extension of quantum mechanics beyond the field of complex
numbers have been studied by different authors \cite{1,2,3,4,5}.
We know from Frobenius that, in the case of algebra without zero
divisor, the investigation must be limited to three fields: real
numbers $\mathbb{R}$, complex numbers $\mathbb{C}$ and quaternions
$\mathbb{H}$. However, recently, some interest have been deployed
to study quantum mechanics beyond the paradigm of algebra without
zero divisors \cite{2,4,7}, principally over hyperbolic numbers
$\mathbb{D}$ (also called duplex numbers in the literature). In
\cite{4} the author has shown that quantum mechanics over the
hyperbolic numbers, called here ``hyperbolic quantum mechanics'',
behaves well 1) in the probabilistic interpretation via the Born's
formula, 2) for the continuity equation $\partial_t
P+\mathbf{\nabla} \cdot \mathbf{J}=0$ ( where $P$ and $\mathbf{J}$
are respectively the scalar-valued ``density'' and vector-valued
``current'') and 3) for the free-particle. However, the main
difference between standard quantum mechanics and hyperbolic
quantum mechanics comes from the fact that they have different
topology on the unit circle. Indeed, the symmetry groups of the
unit circle for complex numbers and hyperbolic numbers are
respectively $SO(2)\sim S^1 \subset \mathbb{C}$ and
$SO_{\uparrow}(1,1)\sim \mathbb{R} \subset \mathbb{D}$. The
consequence of this difference in the topology of the unit circle
is that the superposition of states ``doesn't hold'' in the case
of hyperbolic quantum mechanics. For instance, it is well known
that in the classical Young's two-slit experiment, the intensity
has a sinusoidal pattern. However, in the case of hyperbolic
quantum mechanics, intensity is proportional to the hyperbolic
cosines \cite{4}. Therefore, the fringe pattern cannot be
explained by the hyperbolic quantum mechanics. Nevertheless, in
\cite{2}, it is mentioned that hyperbolic quantum mechanics can be
interesting as a new theory of probability waves that can be
developed in parallel with standard quantum mechanics \cite{7}.

The bicomplex numbers are at the same time a generalization of
complex numbers $\mathbb{C}$ and of hyperbolic numbers
$\mathbb{D}$. Hence, the ``bicomplex quantum mechanics'' is some
generalization of the standard quantum mechanics and of the
hyperbolic quantum mechanics.

In this paper, we investigate the properties of the bicomplex
Schr\"odinger equation. In section~2, we introduce the bicomplex
numbers and present the conjugations and the bicomplex moduli of
these numbers. Then, in section~3, we recall some well known
results on the standard Schr\"odinger equation. In section~4 we
derive the continuity equations, find the discrete symmetries and
introduce the idempotent basis for the bicomplex Schr\"odinger
equation. Finally, in section~5, we introduce the three real
moduli for bicomplex numbers and give bicomplex Born formulas. A
conclusion is made.

\section{Bicomplex numbers}

The bicomplex numbers are defined as \cite{8,9,10}\be
\mathbb{T}:=\{w_0+w_1\bold{i_1}+w_2\bold{i_2}+w_3\bold{j}|\ w_0,
w_1,w_2,w_3 \in \mathbb{R}\}, \ee with the product of imaginary
units given in (\ref{eq:ij}) i.e., $\bz:=1$ acts as identity,
$\bos=\bts=-1$, $\bjs=1$ and \be \ba{rclrcl}
   \bo\bt &=& \bt\bo &=& \bj,  \\
   \bo\bj &=& \bj\bo &=& -\bt, \\
   \bt\bj &=& \bj\bt &=& -\bo. \\
\ea \ee Hence, the bicomplex numbers are commutative. We define
the following two subsets
 $\mathbb{C}(\bold{i_k}) \subset \mathbb{T}$ for $k=1,2$, by
$\mathbb{C}(\bold{i_k}):=\{x+y\bold{i_k}|\bold{i^{\text 2}_k}=-1
\mbox{ and } x,y \in \mathbb{R}\}$.

It is also convenient to write the set of bicomplex numbers as \be
\mathbb{T}=\{z_1+z_2\bold{i_2}|\ z_1,z_2 \in
\mathbb{C}(\bold{i_1})\}. \ee In particular, if we put $z_1=x$ and
$z_2=y\bold{i_1}$ with $x,y \in \mathbb{R}$, then we obtain the
subalgebra of hyperbolic numbers: $\mathbb{D}=\{x+y\bold{j}|\
\bold{j}^2=1, x,y\in \mathbb{R}\}$. (Hyperbolic coordinates are
naturally introduced in special relativity and serve as space-time
coordinates in the Lorentzian's plane, where the non-invertible
coordinates correspond to the light cone, and the elements of the
form $\me^{\phi\bold{j}}$ represent ``boosts'', \cite{11}.)

\subsection{Conjugates for bicomplex numbers}
Complex conjugation plays an important role both for algebraic and
geometric properties of $\mathbb{C}$, as well as in the standard
quantum mechanics. For bicomplex numbers, there are three possible
conjugations. Let $w\in \mathbb{T}$ and $z_1,z_2 \in
\mathbb{C}(\bold{i_1})$ such that $w=z_1+z_2\bold{i_2}$. Then we
define the three conjugations as:

\begin{subequations}
\label{eq:dag}
\begin{align}
w^{\dag_{1}}&=(z_1+z_2\bold{i_2})^{\dag_{1}}:=\overline
z_1+\overline z_2 \bold{i_2},
\\
w^{\dag_{2}}&=(z_1+z_2\bold{i_2})^{\dag_{2}}:=z_1-z_2 \bold{i_2},
\\
w^{\dag_{3}}&=(z_1+z_2\bold{i_2})^{\dag_{3}}:=\overline
z_1-\overline z_2 \bold{i_2},
\end{align}
\end{subequations}
where $\overline z_k$ is the standard complex conjugate of complex
numbers $z_k \in \mathbb{C}(\bold{i_1})$. If we say that the
bicomplex number
$w=z_1+z_2\bold{i_2}=w_0+w_1\bold{i_1}+w_2\bold{i_2}+w_3\bold{j}$
has the ``signature'' $(++++)$, then the conjugations of type 1,2
or 3 of $w$ have, respectively, the signatures $(+-+-)$, $(++--)$
and $(+--+)$.

We can verify easily that each of these conjugates can be
expressed in terms of the two others, i.e.
$w^{\dag_{3}}=(w^{\dag_{1}})^{\dag_{2}}=(w^{\dag_{2}})^{\dag_{1}}$,
etc. More precisely, under the composition, the conjugates form
the following abelian group:
\begin{center}
\be
\begin{tabular}{|c||c|c|c|c|}
\hline
$\circ$ & $\dag_{0}$ & $\dag_{1}$  & $\dag_{2}$  & $\dag_{3}$  \\
\hline
\hline
$\dag_{0}$    & $\dag_{0}$ & $\dag_{1}$  & $\dag_{2}$  & $\dag_{3}$  \\
\hline
$\dag_{1}$    & $\dag_{1}$ & $\dag_{0}$ & $\dag_{3}$  & $\dag_{2}$ \\
\hline
$\dag_{2}$    & $\dag_{2}$ & $\dag_{3}$  & $\dag_{0}$ & $\dag_{1}$ \\
\hline
$\dag_{3}$    & $\dag_{3}$ & $\dag_{2}$ & $\dag_{1}$ & $\dag_{0}$ \\
\hline
\end{tabular}
\label{eq:groupedag} \ee
\end{center}
where $w^{\dag_{0}}:=w\mbox{ } \forall w\in \mathbb{T}$.

The three kind of conjugations all have the standard properties of
conjugations, i.e.
\begin{eqnarray}
(s+ t)^{\dag_{k}}&=&s^{\dag_{k}}+ t^{\dag_{k}},\\
\left(s^{\dag_{k}}\right)^{\dag_{k}}&=&s, \\
\left(s\cdot t\right)^{\dag_{k}}&=&s^{\dag_{k}}\cdot t^{\dag_{k}}.
\end{eqnarray}
for $s,t \in \mathbb{T}$ and $k=0,1,2,3$. The proofs of these
properties are rather technical and simple. Nevertheless, let us
illustrate the proof for the last property in the case of the
conjugation of the first kind. Let $s=z_1+z_2\bold{i_2}$ and
$t=z_3+z_4\bold{i_2}$ with $z_1,z_2,z_3,z_4 \in
\mathbb{C}(\bold{i_1})$, then
\[
\ba{rcl} \left(s\cdot t\right)^{\dag_{1}}&=&
\left[(z_1z_3-z_2z_4)+(z_1z_4+z_2z_3)\bold{i_2}\right]^{\dag_{1}}\\*[2ex]
&=&
\overline{(z_1z_3-z_2z_4)}+\overline{(z_1z_4+z_2z_3)}\bold{i_2}
\\*[2ex]
&=&
(\overline{z_1z_3}-\overline{z_2z_4})+(\overline{z_1z_4}+\overline{z_2z_3})\bold{i_2}
\\*[2ex]
&=&(\overline z_1+\overline z_2 \bold{i_2})\cdot(\overline
z_3+\overline z_4 \bold{i_2})\\*[2ex] &=& s^{\dag_{1}}\cdot
t^{\dag_{1}}. \ea
\]

\subsection{The bicomplex moduli}
We know that the product of a standard complex number with his
conjugate gives the square of the Euclidean metric in
$\mathbb{R}^2$. The analog of these, for bicomplex numbers, are
the following. Let $z_1,z_2 \in \mathbb{C}(\bold{i_1})$ and
$w=z_1+z_2\bold{i_2}\in \mathbb{T}$, then we have that \cite{10}:
\begin{subequations}
\begin{align}
|w|^{2}_{\bold{i_1}}&:=w\cdot w^{\dag_{2}}=z^{2}_{1}+z^{2}_{2} \in
\mathbb{C}(\bold{i_1}),
\\*[2ex] |w|^{2}_{\bold{i_2}}&:=w\cdot
w^{\dag_{1}}=\left(|z_1|^2-|z_2|^2\right)+2\pre(z_1\overline
z_2)\bold{i_2} \in \mathbb{C}(\bold{i_2}), \\*[2ex]
|w|^{2}_{\bold{j}}&:=w\cdot
w^{\dag_{3}}=\left(|z_1|^2+|z_2|^2\right)-2\pim(z_1\overline
z_2)\bold{j} \in \mathbb{D},
\end{align}
\end{subequations}
where the subscript of the square modulus refers to the subalgebra
$\mathbb{C}(\bold{i_1}), \mathbb{C}(\bold{i_2})$ or $\mathbb{D}$
of $\mathbb{T}$ in which $w$ is projected.
Note that for $z_1,z_2 \in \mathbb{C}(\bold{i_1})$ and
$w=z_1+z_2\bold{i_2}\in \mathbb{T}$, we can define the usual norm
of $w$ as
$|w|=\sqrt{|z_1|^2+|z_2|^2}=\sqrt{\pre(|w|^{2}_{\bold{j}})}$.

It is easy to verify that $w\cdot \displaystyle
\frac{w^{\dag_{2}}}{|w|^{2}_{\bold{i_1}}}=1$. Hence, the inverse
of $w$ is given by \be w^{-1}= \displaystyle
\frac{w^{\dag_{2}}}{|w|^{2}_{\bold{i_1}}}. \ee From this, we find
that the set $\mathcal{NC}$ of zero divisors of $\mathbb{T}$,
called the {\em null-cone}, is given by $\{z_1+z_2\bold{i_2}|\
z_{1}^{2}+z_{2}^{2}=0\}$, which can be rewritten as \be
\mathcal{NC}=\{z(\bold{i_1}\pm\bold{i_2})|\ z\in
\mathbb{C}(\bold{i_1})\}. \ee

\subsection{Exponential function}
Contrary to quaternions, the exponential function is well defined
on bicomplex numbers and posses all the standard properties.
Hence, for $z_1,z_2 \in \mathbb{C}(\bold{i_1})$ and
$w=z_1+z_2\bold{i_2}\in \mathbb{T}$, we have  \be
\me^{w}=\me^{z_1+z_2\bold{i_2}}=\me^{z_1}\me^{z_2\bold{i_2}}=\me^{z_1}(\cos
z_2+\bold{i_2}\sin z_2), \label{eq:e}\ee corresponding to {\em
hyper-polar} coordinates. It is easy to see that this is a
generalization of the polar coordinates for the complex {\em and}
the hyperbolic numbers. Indeed, in particular for $z_1=\ln r \in
\mathbb{R}$ and $z_2=\theta \in \mathbb{R}$, we obtain the
standard complex polar coordinates. If $z_1=\ln \rho\in
\mathbb{R}$ and $z_2=\phi\bold{i_1}$ with $\phi \in \mathbb{R}$,
then the equation (\ref{eq:e}) becomes \be \ba{rcl}
\me^{z_1+z_2\bold{i_2}} =  \rho e^{\phi\bold{j}}&=&
\rho\left[\cos( \phi\bold{i_1})+\bold{i_2}\sin(\phi\bold{i_1})\right]\\
&=& \rho\left[\cosh\phi+\bold{i_2}\bold{i_1}\sinh \phi \right]\\
&=& \rho\left[\cosh\phi+\bold{j} \sinh \phi\right], \ea \ee which
corresponds to the hyperbolic polar coordinates used in references
\cite{2,4,7,9,11,12}.

As in the standard case, the bicomplex number $\me^{w}$,
is always invertible $\forall w\in \mathbb{T}$. Moreover, we have these
 useful properties for all the conjugates:
\be (\me^{w})^{\dag_k}=(\me^{w^{\dag_k}})\mbox{, } k=0,1,2,3. \ee

\section{The standard Schr\"odinger's equation} Before going in
the analysis of the bicomplex Schr\"odinger equation, let us first
review some well known results of the standard one-dimensional
Schr\"odinger's equation:
\begin{equation}
\bold{i}\hbar\,\partial_t\psi(x,t)+\frac{\hbar^2}{2m}\,
\partial^{2}_{x}\psi(x,t)-V(x,t)\psi(x,t)=0
\label{eq:s}
\end{equation}
where
\[\psi:\mathbb{R}^2\rightarrow \mathbb{C} \mbox{ and } V:
\mathbb{R}^2\rightarrow \mathbb{R}.\] First, if we set
$\psi=\me^{\alpha(x,t)+\beta(x,t)\bold{i}}$ with
$\alpha,\beta:\mathbb{R}^2\rightarrow \mathbb{R}$, then it is well
known that the Schr\"odinger's equation can be rewritten in a
system of two differential equations in terms of the functions
$\alpha$ and $\beta$:
\begin{subequations}
\label{eq:ess}
\begin{align}
-\hbar\, \partial_t\beta&+\frac{\hbar^2}{2m}\left[
\partial^{2}_{x}\alpha+(\partial_{x}\alpha)^2-(\partial_x\beta)^2\right]-V=0,
\label{eq:1ess}
\\
\partial_t\alpha&+\frac{\hbar}{2m}\left[
\partial^{2}_{x}\beta+2\, \partial_x\alpha\,
\partial_x\beta\right]=0.
\label{eq:2ess}
\end{align}
\end{subequations}
The probability density $P(\psi)$ to find a particle in the state
$\psi(x,t)$ is then given by the Born's formula: \be
P(\psi)=\psi\overline\psi=\me^{2\al}. \label{eq:Prob} \ee

One other very well known result of the standard Schr\"odinger's
equation is the conservation of the probability current
\be\ba{lrcl} &\partial_t(\psi\overline\psi)&+&\nabla\cdot
\mathbf{J(\psi)}=0, \label{eq:consj} \ea\ee
\hspace{3cm}\mbox{where}
\begin{subequations}
\begin{align}
J(\psi)&=\displaystyle
\frac{\hbar}{2m\bold{i}}(\overline\psi\partial_x\psi-\psi\partial_x\overline\psi)\,
\label{eq:consjj} \\*[2ex] &=
\displaystyle\frac{\hbar}{m}\me^{2\alpha}\partial_x \beta\,.
\end{align}
\end{subequations}
We note that equation (\ref{eq:2ess}) and the conservation of the
current probability (\ref{eq:consj}) coincide. Hence, by
decomposing the standard Schr\"odinger's equation into his real
and imaginary parts, one obtain two equations: (\ref{eq:1ess})
corresponding to an extended version of the Jacobi-Hamilton
equation and (\ref{eq:2ess}) corresponding to the conservation of
the probability current.

\section{The bicomplex Schr\"odinger equation}

Let us now consider an analog of the one-dimensional standard
Schr\"odinger's equation over the bicomplex space functions: \be
\bold{i_1}\hbar\, \partial_t\psi(x,t)+\frac{\hbar^2}{2m}\,
\partial^{2}_{x}\psi(x,t)-V(x,t)\psi(x,t)=0
\label{eq:bs}\ee
where
\[\psi:\mathbb{R}^2\rightarrow \mathbb{T}
\mbox{ and } V: \mathbb{R}^2\rightarrow \mathbb{R}.\]

The choice of the imaginary unit $\bold{i_1}$ appearing explicitly
in the bicomplex Schr\"odinger equation is not arbitrary. In the
case of hyperbolic quantum mechanics it have been shown that the
choice of the hyperbolic imaginary unit $\bold{j}$ doesn't yield
the superposition principle \cite{4}. Hence, in our case, we
should choose between the imaginary units $\bold{i_1}$ or
$\bold{i_2}$. In principle there is no major raison to prefer
$\bold{i_1}$ instead of $\bold{i_2}$, however we will see later
that imaginary unit $\bold{i_1}$ is more appropriated for the
decomposition of the bicomplex Schr\"odinger equation into what we
will call the \textit{idempotent basis}.

We express the wave function $\psi(x,t)$ into the {\em
hyper-polar} coordinates as \be
\psi(x,t)=\me^{z_1(x,t)+z_2(x,t)\bold{i_2}}, \label{eq:fo}\ee
where
\begin{subequations}
\begin{align}
z_1(x,t)&=\alpha(x,t)+\beta(x,t)\bold{i_1}, \\*[2ex]
z_2(x,t)&=\gamma(x,t)+\delta(x,t)\bold{i_1},
\end{align}
\end{subequations}
and $\al,\b,\ga$ and $\de$ are real functions going from
$\mathbb{R}^2\rightarrow \mathbb{R}$. Hence, one can decompose the
bicomplex Schr\"odinger equation (\ref{eq:bs}) into a system of
four differential equations in terms of the four real functions
$\alpha,\beta,\gamma$ and $\delta$:

\begin{subequations}
\label{eq:0}
\begin{align}
-\hbar\, \partial_t\beta &+ \frac{\hbar^2}{2m}\left[
\partial^{2}_{x}\alpha + (\partial_{x}\alpha)^2-(\partial_x\beta)^2-(\partial_x\gamma)^2+(\partial_x\delta)^2\right]-V=0,
\label{eq:1}
\\
\partial_t\alpha &+ \frac{\hbar}{2m}\left[
\partial^{2}_{x}\beta+2(\partial_x\alpha\,\partial_x\beta-\partial_x\gamma\,\partial_x\delta)\right]=0,
\label{eq:2}
\\
-\partial_t\delta &+ \frac{\hbar}{2m}\left[
\partial^{2}_{x}\gamma+2(\partial_x\alpha\partial_x\gamma-\partial_x\beta\partial_x\delta)\right]=0,
\label{eq:3}
\\
\partial_t\gamma &+ \frac{\hbar}{2m}\left[
\partial^{2}_{x}\delta+2(\partial_x\alpha\,\partial_x\delta+\partial_x\beta\,\partial_x\gamma)\right]=0.
\label{eq:4}
\end{align}
\end{subequations}
We remark that when $\gamma\rightarrow 0$ and $\delta\rightarrow
0$ the system of equations (\ref{eq:0}) goes to the system
(\ref{eq:ess}) of the standard Schr\"odinger's equation.

\subsection{The bicomplex continuity equations}
In this section, we derive the continuity equations for the
bicomplex Schr
\"odinger equation. For that, we rewrite the
bicomplex Schr\"odinger equation under the four kind of
conjugations:
\begin{subequations}
\label{eq:esd}
\begin{align}
\bold{i_1}\hbar\,
\partial_t\psi+\frac{\hbar^2}{2m}\,\partial^{2}_{x}\psi-V\psi&=0,
\label{eq:esd0}
\\*[2ex]
-\bold{i_1}\hbar\,
\partial_t\psi^{\dag_1}+\frac{\hbar^2}{2m}\,\partial^{2}_{x}\psi^{\dag_1}-V\psi^{\dag_1}&=0,
\label{eq:esd1}
\\*[2ex]
\bold{i_1}\hbar\,
\partial_t\psi^{\dag_2}+\frac{\hbar^2}{2m}\,\partial^{2}_{x}\psi^{\dag_2}-V\psi^{\dag_2}&=0,
\label{eq:esd2}
\\*[2ex]
-\bold{i_1}\hbar\,
\partial_t\psi^{\dag_3}+\frac{\hbar^2}{2m}\,\partial^{3}_{x}\psi^{\dag_3}-V\psi^{\dag_3}&=0.
\label{eq:esd3}
\end{align}
\end{subequations}
Let us first consider equations (\ref{eq:esd0}) and
(\ref{eq:esd1}). Multiplying (\ref{eq:esd0}) by $\psi^{\dag_1}$
and (\ref{eq:esd1}) by $\psi$, and subtracting these two
equations, one obtain \be
\bold{i_1}\partial_t(\psi\psi^{\dag_1})+\frac{\hbar}{2m}(\psi^{\dag_1}\partial^{2}_{x}\psi-\psi\partial^{2}_{x}\psi^{\dag_1})=0,
\ee which can be rewritten into the continuity equation
\be\ba{lrcl} &\partial_t(\psi \psi^{\dag_1})&+&\nabla\cdot
\mathbf{J}_1(\psi)=0, \label{eq:consj1}

\ea \ee \hspace{2.5cm}\mbox{where}
\begin{subequations}
\begin{align}
J_1(\psi)&=\displaystyle
\frac{\hbar}{2m\bold{i_1}}(\psi^{\dag_1}\partial_x\psi-\psi\partial_x
\psi^{\dag_1}),
\\*[2ex] &=
\displaystyle\frac{\hbar}{m}\me^{2(\alpha+\gamma\bold{i_2})}\partial_x(
\beta+\delta\bold{i_2})\,. \label{eq:consj1.1}
\end{align}
\end{subequations}
These calculations can be done for all pair of equations in the
set of equations (\ref{eq:esd}). However, one can construct a
continuity equation only if the sign of the imaginary unit
$\bold{i_1}$, in front of each equation, are opposite into the
pair of equations that we consider. Indeed, it is not possible to
obtain a continuity equation from the pair of equations
(\ref{eq:esd0}) and (\ref{eq:esd2}) or (\ref{eq:esd1}) and
(\ref{eq:esd3}). Therefore, we find three other continuity
equations (for a total of four): \be \partial_t(\psi
\psi^{\dag_3})+\nabla\cdot \mathbf{J}_2(\psi)=0, \label{eq:consj2}
\ee
\begin{subequations}
\begin{align}
J_2(\psi)&=\displaystyle
\frac{\hbar}{2m\bold{i_1}}(\psi^{\dag_3}\partial_x\psi-\psi\partial_x
\psi^{\dag_3}),
\\*[2ex] &=
\displaystyle\frac{\hbar}{m}\me^{2(\alpha+\delta\bold{j})}\partial_x(
\beta-\gamma\bold{j})\,. \label{eq:consj2.1}
\end{align}
\end{subequations}

\be \partial_t(\psi^{\dag_2} \psi^{\dag_1})+\nabla\cdot
\mathbf{J}_3(\psi)=0,\hspace{4cm} \label{eq:consj3} \ee
\begin{subequations}
\begin{align}
J_3(\psi)&=\displaystyle
\frac{\hbar}{2m\bold{i_1}}(\psi^{\dag_1}\partial_x\psi^{\dag_2}-\psi^{\dag_2}\partial_x
\psi^{\dag_1}),
\\*[2ex] &=
\displaystyle\frac{\hbar}{m}\me^{2(\alpha-\delta\bold{j})}\partial_x(
\beta+\gamma\bold{j}) \,. \label{eq:consj3.1}
\end{align}
\end{subequations}

\be \partial_t(\psi^{\dag_2} \psi^{\dag_3})+\nabla\cdot
\mathbf{J}_4(\psi)=0,\hspace{4cm} \label{eq:consj4} \ee
\begin{subequations}
\begin{align}
J_4(\psi)&=\displaystyle
\frac{\hbar}{2m\bold{i_1}}(\psi^{\dag_3}\partial_x\psi^{\dag_2}-\psi^{\dag_2}\partial_x
\psi^{\dag_3}),
\\*[2ex] &=
\displaystyle\frac{\hbar}{m}\me^{2(\alpha-\gamma\bold{i_2})}\partial_x(
\beta-\delta\bold{i_2}) \,. \label{eq:consj4.1}
\end{align}
\end{subequations} However, on can verify that these four
continuity equations are an over determined system of equations.
Indeed, if we conjugate equation (\ref{eq:consj1}) by $\dag_2$, we
obtain equation (\ref{eq:consj4}) where
$\mathbf{J}_4=(\mathbf{J}_1)^{\dag_2}$. In the same way, if we
conjugate equation (\ref{eq:consj2}) by $\dag_1$ we obtain
equation (\ref{eq:consj3}) with
$\mathbf{J}_3=(\mathbf{J}_2)^{\dag_1}$. Therefore, we have two
independent continuity equations. Let us choose $\mathbf{J}_1$ and
$\mathbf{J}_2$ as the ``basis'' for the currents.

The continuity equations (\ref{eq:consj1}) is in fact equivalent
to the equations (\ref{eq:2}) and (\ref{eq:4}). In the same way,
the continuity equations (\ref{eq:consj2}) is equivalent to the
equations (\ref{eq:2}) and (\ref{eq:3}). Therefore, equation
(\ref{eq:1}) corresponds to an extended version of the
Hamilton-Jacobi equation of the standard case and the system
(\ref{eq:2}), (\ref{eq:3}) and (\ref{eq:4}) are equivalent to the
continuity equations (\ref{eq:consj1}) and (\ref{eq:consj2})
expresed in terms of $\mathbf{J}_1$ and $\mathbf{J}_2$ only.

\subsection{Discrete symmetries of the bicomplex Schr\"odinger \\equation}
The system of equations (\ref{eq:0}) posses a 8-dimensional
discrete group, leaving the solution set of the system invariant.
These discrete symmetry group is given by \be \ba{rclrcl} \hat
P_0&=&Id & \hat P_1&=&\left\{ \ba{rcr} \gamma&\rightarrow& -\gamma
\\ \delta&\rightarrow& -\delta \ea \right.
\\*[2ex]
\hat P_2&=&\left\{ \ba{rcr}
\gamma&\rightarrow& -\delta \bold{i_2}  \\
\delta&\rightarrow& \gamma \bold{i_2}  \ea \right. & \hat
P_3&=&\left\{ \ba{rcr}
\gamma&\rightarrow& \delta \bold{i_2}  \\
\delta&\rightarrow& -\gamma \bold{i_2}  \ea \right.
\\*[2ex]
\hat P_4&=&\left\{ \ba{rcr}
\gamma&\rightarrow& \delta \bold{i_1}  \\
\delta&\rightarrow& -\gamma \bold{i_1}  \ea \right. & \hat
P_5&=&\left\{ \ba{rcr}
\gamma&\rightarrow& -\delta \bold{i_1}  \\
\delta&\rightarrow& \gamma \bold{i_1}  \ea \right.
\\*[2ex]
\hat P_6&=&\left\{ \ba{rcr}
\gamma&\rightarrow& \gamma \bold{j}  \\
\delta&\rightarrow& \delta \bold{j}  \ea \right. & \hat
P_7&=&\left\{ \ba{rcr}
\gamma&\rightarrow& -\gamma \bold{j}  \\
\delta&\rightarrow& -\delta \bold{j}  \ea \right.. \ea
\label{eq:groupe}
 \ee Note that functions $\alpha(x,t)$ and $\beta(x,t)$, of the
bicomplex wave function $\psi(x,t)$, are not transformed under
these discrete symmetry group.

Let us mention some remarks about these transformations. First,
the group of symmetry (\ref{eq:groupe}) is an abelian group with
$\hat P^2_n=Id$ for $n=0,1,\ldots,7$. Second, the set given by
$\{\hat P_0,\hat P_1,\hat P_2,\hat P_3\}$ is a subgroup of the
symmetry group and is isomorphic to the group of conjugates
(\ref{eq:groupedag}) for the bicomplex numbers. Finally, we remark
that the discrete operators $\hat P_4,\hat P_5,\hat P_6$ and $\hat
P_7$ act on an arbitrary bicomplex number $w$ exactly as the
discrete operators $\hat P_0,\hat P_1,\hat P_2$ and $\hat P_3$,
respectively, i.e. \be \hat P_{n+4}(w)=\hat P_n(w), \mbox{ for }
n=0,1,2,3\ \forall w\in \mathbb{T}. \ee Hence we have in fact a
``fundamental subgroup'', for the symmetry group, given by $\{\hat
P_0,\hat P_1,\hat P_2,\hat P_3\}$.

Let us now apply these symmetries on the system of equations
equivalent to (\ref{eq:0}), i.e. the system of equations
consisting of (\ref{eq:1}) and the two continuity equations
(\ref{eq:consj1}) and (\ref{eq:consj2}). We already know that
equation (\ref{eq:1}) is invariant under the symmetries. Let us
now look how the continuity equations (\ref{eq:consj1}) and
(\ref{eq:consj2}) are transformed under these symmetries. For
that, we only have to calculate the action of the symmetry
operators on $\psi$ (since $\mathbf J_1$ and $\mathbf J_2$ are
expressible in term of $\psi$). We find that \be \ba{l}

\hat P_1(\psi)=\psi^{\dag_2},\ \hat P_2(\psi)=\psi_+,\ \hat
P_3(\psi)=\psi_-,

 \ea \ee
where the functions $\psi_+$ and $\psi_-$ are functions in the
$\mathbb{C}(\bold{i_1})$-space given by \be \psi_{\pm}=\me^{z_1\mp
z_2\bold{i_1}}=\me^{(\alpha\pm\delta)+(\beta\mp
\gamma)\bold{i_1}}. \label{eq:fopm}
 \ee From these calculations, we find that $\hat P_1$ transforms
equations (\ref{eq:consj1}) and (\ref{eq:consj2}) into equations
(\ref{eq:consj4}) and (\ref{eq:consj3}), respectively. Under the
discrete symmetry $\hat P_2$, the equations (\ref{eq:consj1}) and
(\ref{eq:consj2}) are both transformed into the continuity
equation

\be
\partial_t(\psi_+ \overline \psi_+)+\nabla\cdot
\mathbf{J}(\psi_+)=0 \label{eq:ecj+}, \ee where
\begin{subequations}
\label{eq:consj+}
\begin{align}
J(\psi_+)&=\displaystyle
\frac{\hbar}{2m\bold{i_1}}(\overline\psi_+\partial_x\psi_+-\psi_+\partial_x\overline\psi_+)\,
 \\*[2ex]
&=\frac{\hbar}{m}\me^{2(\alpha+\delta)}
\partial_x(\beta-\gamma).
\end{align}
\end{subequations}
 Finally, under $\hat P_3$, the equations (\ref{eq:consj1}) and
(\ref{eq:consj2}) are both transformed into the continuity
equation \be
\partial_t(\psi_- \overline \psi_-)+\nabla\cdot
\mathbf{J}(\psi_-)=0, \label{eq:ecj-} \ee
\begin{subequations}
\label{eq:consj-}
\begin{align}
J(\psi_-)&=\displaystyle
\frac{\hbar}{2m\bold{i_1}}(\overline\psi_-\partial_x\psi_--\psi_-\partial_x\overline\psi_-)\,
 \\*[2ex] &=\frac{\hbar}{m}\me^{2(\alpha-\delta)}
\partial_x(\beta+\gamma).
\end{align}
\end{subequations}

Hence, under the symmetry operators, we have recover the two
continuity equations (\ref{eq:consj4}) and (\ref{eq:consj3}),
dropped previously since respectively equivalent to
(\ref{eq:consj1}) and (\ref{eq:consj2}). Moreover, we have found
two new continuity equations (\ref{eq:ecj+}) and (\ref{eq:ecj-}),
associated with two {\em real} currents $\mathbf{J}(\psi_+)$ and
$\mathbf{J}(\psi_-)$.

Note that it is possible to express the bicomplex wave function
$\psi$ in terms of $\psi_+$ and $\psi_-$ by using what is called
the {\em idempotents basis}. Indeed, for all bicomplex numbers,
one can pass from the real basis
$\{1,\bold{i_1},\bold{i_2},\bold{j}\}$ to the complex (in
$\bold{i_1}$) basis $\{\bold{e_1} , \bold{e_2}\}$, where
$\bold{e_1}=\frac{1+\bold{j}}{2}$,
$\bold{e_2}=\frac{1-\bold{j}}{2}$ (in fact $\bold{e_2}$ can be
rewritten in term of $\bold{e_1}$, i.e.
$\bold{e_2}=(\bold{e_1})^{\dag_1}=(\bold{e_1})^{\dag_2}$, but
$\bold{e_2}\not= (\bold{e_1})^{\dag_3}$). The elements
$\bold{e_1}, \bold{e_2}$ having the following properties: \be
(\bold{e_1})^2=\bold{e_1},\ (\bold{e_2})^2=\bold{e_2},\
\bold{e_1}\bold{e_2}=0. \ee Every bicomplex numbers
$z_1+z_2\bold{i_2}$, $z_1,z_2 \in \mathbb{C}(\bold{i_1})$, can be
expressed in the idempotent basis as \be
z_1+z_2\bold{i_2}=(z_1-z_2\bold{i_1})\bold{e_1}+(z_1+z_2\bold{i_1})\bold{e_2}.
\label{eq:id}\ee Moreover, the bicomplex exponential can be
rewritten as follows \be
\me^{z_1+z_2\bold{i_2}}=\me^{z_1-z_2\bold{i_1}}\bold{e_1}
+\me^{z_1+z_2\bold{i_1}}\bold{e_2}. \ee

In the same way, we can express the wave function in the
idempotent basis as \be \psi=\psi_+\, \bold{e_1} +\psi_-\,
\bold{e_2}. \label{eq:psiid} \ee Hence, using the idempotent
basis, we can rewrite the bicomplex Schr\"odinger equation
(\ref{eq:bs}) in the form \[ \left(\bold{i_1}\hbar
\partial_t\psi_{+} + \displaystyle\frac{\hbar^2}{2m}
\partial^{2}_{x}\psi_{+}-V\psi_{+}\right)\bold{e_1}
+\left(\bold{i_1}\hbar \partial_t\psi_{-} +
\displaystyle\frac{\hbar^2}{2m}
\partial^{2}_{x}\psi_{-}-V\psi_{-}\right)\bold{e_2}=0,
\] which can be decomposed into the following two standard
Schr\"odinger's equations (complex in $\bold{i_1}$):

\be \bold{i_1}\hbar\,\partial_t\psi_{\pm}+\displaystyle
\frac{\hbar^2}{2m}\,
\partial^{2}_{x}\psi_{\pm}-V\psi_{\pm}=0. \ee
Associated with these equations, it is now obvious to see that we
have the continuity equations (\ref{eq:ecj+}) and (\ref{eq:ecj-})
written, respectively, in terms of the real currents
$J(\psi_{\pm})$ given by (\ref{eq:consj+}) and (\ref{eq:consj-}).

\section{The bicomplex Born formula}
In the case of the standard Schr\"odinger's equation (linear and
homogeneous) it is well known that the continuous symmetries of
(\ref{eq:s}), acting on a solution $\psi$ only, are \be \ba{rcl}
\psi &\rightarrow& \lambda\psi,\ \lambda \in \mathbb{C}, \\
\psi &\rightarrow& \psi+\phi\ (\phi\mbox{ an other solution of the
Schr\"odinger's eq.}),
 \ea \ee corresponding, respectively, to a
dilation of the wave function and  the superposition principle. In
quantum mechanics dilation is used for the normalization of the
wave function and the superposition principle is one fundamental
property.

In particular, the dilation can be expressed as a rotation of the
wave function when $\lambda=\me^{i\theta}$. These particular
symmetry plays an important role in quantum mechanics since it is
invariant under the Born's formula.

For the bicomplex Schr\"odinger equation we still have the
dilation (with $\lambda \in \mathbb{T}$) and the superposition
principle (where $\psi$ and $\phi$ are bicomplex functions) as
continuous symmetries, since equation (\ref{eq:bs}) is linear and
homogeneous. However, in addition, we have the discrete symmetries
given in (\ref{eq:groupe}). In this section, we want to study the
bicomplex discrete symmetries for bicomplex Born formulas.

\subsection{Definitions of the real moduli}
In order to obtain some bicomplex Born formula from our bicomplex
wave function $\psi(x,t)$, let us define the following three
\textit{real moduli} (see \cite{10}):

\begin{enumerate}
\item[\textbf{1)}] For $s,t\in \mathbb{T}$, we define the first
modulus as $|\cdot|_{\bold{1}}:=||\cdot|_{\bold{i_1}}|$. This
modulus has the following properties:
\begin{enumerate}
\item[a)] $|\cdot|_{\bold{1}}: \mathbb{T}\rightarrow \mathbb{R}$,
\item[b)] $|s|_{\bold{1}}\ge 0$ with $|s|_{\bold{1}}=0$ iff $s\in
\mathcal{NC}$, \item[c)] $|s\cdot
t|_{\bold{1}}=|s|_{\bold{1}}\cdot |t|_{\bold{1}}$.
\end{enumerate}From this definition, we
can rewrite this real pseudo-modulus in a much practical point of
view as \be
|w|_{\bold{1}}=|z^{2}_{1}+z^{2}_{2}|^{1/2},\label{eq:wr1.1} \ee
for $w=z_1+z_2\bold{i_2}$ with $z_1,z_2 \in
\mathbb{C}(\bold{i_1})$. Moreover, it is also useful to express
$|\cdot|_{\bold{1}}$, in terms of our three bicomplex conjugates,
i.e.
\be |w|_{\bold{1}}=\sqrt[4]{ww^{\dag_1}w^{\dag_2}w^{\dag_3}}. \label{eq:wr1.2} \ee

\item[\textbf{2)}] For $s,t\in \mathbb{T}$, we define the second
modulus as $|\cdot|_{\bold{2}}:=||\cdot|_{\bold{i_2}}|$. This
modulus has the same properties as $|\cdot|_{\bold{1}}$. Indeed we
can rewrite $|w|_{\bold{2}}$ as \be
|w|_{\bold{2}}=|z^{2}_{1}+z^{2}_{2}|^{1/2},\ee where
$w=z_1+z_2\bold{i_1}$ with $z_1,z_2 \in \mathbb{C}(\bold{i_2})$.
Hence, the first and the second pseudo-modulus are the same.

\item[\textbf{3)}] For $s,t\in \mathbb{T}$, we define the third
modulus as $|\cdot|_{\bold{3}}:=||\cdot|_{\bold{j}}|$. This
modulus has the following properties:
\begin{enumerate}
\item[a)] $|\cdot|_{\bold{3}}: \mathbb{T}\rightarrow \mathbb{R}$,
\item[b)] $|s|_{\bold{3}}\ge 0$ with $|s|_{\bold{3}}=0$ iff $s=0$,
\item[c)] $|s+t|_{\bold{3}}\leq |s|_{\bold{3}}+|t|_{\bold{3}}$,
\item[d)] $|s\cdot t|_{\bold{3}}\leq\sqrt{2}|s|_{\bold{3}}\cdot
|t|_{\bold{3}}$.
\end{enumerate} We note that
\begin{enumerate}
\item[(i)]
$|w|_{\bold{j}}=|z_1-z_2\bold{i_1}|\bold{e_1}+|z_1+z_2\bold{i_1}|\bold{e_2}\in\mathbb{D}$
$\forall w=z_1+z_2\bold{i_2}\in\mathbb{T}$, \item[(ii)] $|s\cdot
t|_{\bold{j}}=|s|_{\bold{j}}|t|_{\bold{j}}\mbox{ }\forall
s,t\in\mathbb{T}$.
\end{enumerate}
\noindent From this definition, we can rewrite the modulus
$|\cdot|_{\bold{3}}$ as \be |w|_{\bold{3}}=\sqrt{|z_1|^2+|z_2|^2},
\label{eq:wr3.1} \ee for $w=z_1+z_2\bold{i_2}$ with $z_1,z_2 \in
\mathbb{C}(\bold{i_1})$. Hence, we see that in fact
$|\cdot|_{\bold{3}}$ is simply the Euclidean metric of
$\mathbb{R}^{4}$, i.e. \be
|w|_{\bold{3}}=|w|=\sqrt{\pre(|w|^{2}_{\bold{j}})}.
\ee
\end{enumerate}

\subsection{Invariance under the discrete symmetries for the real moduli}
Let us first calculate the \textit{real moduli} for the bicomplex
wave function $\psi(x,t)$. We obtain the following ``bicomplex
Born formulas'':

 \bea |\psi|^{2}_{\bold{1}}&=& |\psi|^{2}_{\bold{2}}= \me^{2\alpha},
\\*[2ex]
|\psi|^{2}_{\bold{3}}&=&\me^{2\alpha}\cosh(2\delta)=\me^{2\alpha}\left(1+\displaystyle
\frac{(2\delta)^2}{2!}+\displaystyle
\frac{(2\delta)^4}{4!}+\cdots\right).
 \eea
For $|\psi|^{2}_{\bold{1}}$, we find the same result as in the
standard case (\ref{eq:Prob}) and $|\psi|^{2}_{\bold{3}}$ is some
kind of {\em hyperbolic perturbation} of the standard case when
$\delta(x,t)$ is small.

Let us now consider the invariance of $|\psi|^{2}_{\bold{k}}$
($k=1,2,3$) under the discrete symmetries. To illustrate that, we
first consider the operator $\hat P_1$. A new wave function
$\tilde{\psi}$ is obtained by applying the symmetry operator $\hat
P_1$ on $\psi$. Then by calculating the result on
$|\tilde{\psi}|^{2}_{\bold{1}}$, we find \be
\ba{rcl}|\tilde{\psi}|^{2}_{\bold{1}}:=|\hat
P_1\psi|^{2}_{\bold{1}}&=& \sqrt{(\hat P_1\psi)(\hat
P_1\psi)^{\dag_1}(\hat P_1\psi)^{\dag_2}(\hat P_1\psi)^{\dag_3}}
\\*[2ex] &=&\sqrt{\psi^{\dag_2}\psi^{\dag_3}\psi \psi^{\dag_1}}
\\*[2ex] &=& |\psi|^{2}_{\bold{1}}=\me^{2\alpha}.\ea \ee
Therefore, $|\psi|^{2}_{\bold{1}}$ is invariant under $\hat P_1$.
Performing these calculations for all the \textit{real moduli}, under all
the discrete symmetries, we obtain
\begin{equation}
|\hat P_1\psi|^{2}_{\bold{k}} =|\psi|^{2}_{\bold{k}}=
\begin{cases}
\me^{2\alpha} &  \text{if $k=1,2$}\\
\me^{2\alpha}\cosh(2\delta) & \text{if $k=3$}
\end{cases}
\end{equation}
and
\be \ba{rcl}
|\hat
P_2\psi|^{2}_{\bold{k}}&=&\me^{2\delta}|\psi|^{2}_{\bold{k}}=\me^{2(\alpha+\delta)},
\\*[2ex]
|\hat
P_3\psi|^{2}_{\bold{k}}&=&\me^{-2\delta}|\psi|^{2}_{\bold{k}}=\me^{2(\alpha-\delta)}
 \ea \ee
for $k=1,2,3$.

It is now easy to prove that the bicomplex Born formulas will be
preserved under {\em all} our discrete symmetries
(\ref{eq:groupe}) if and only if the wave function $\psi$ has the
form
\be\psi(x,t)=\me^{\alpha(x,t)+\beta(x,t)\bold{i_1}+\gamma(x,t)\bold{i_2}},\ee
i.e. if and only if $\delta(x,t)=0$. Moreover, we have the
following result for $|\cdot|^{2}_{\bold{3}}$:

\begin{theorem}
Let $\psi$ be a bicomplex wave function given by $\psi(x,t)=\\
\me^{\alpha(x,t)+\beta(x,t)\bold{i_1}+\gamma(x,t)\bold{i_2}+\delta(x,t)\bold{j}}= \psi_{+}(x,t)\, \bold{e_1}+\psi_{-}(x,t)\, \bold{e_2}$.
Then, \be |\psi(x,t)|^{2}_{\bold{3}}=\frac{|\psi_+|^{2}+|\psi_-|^{2}}{2}. \ee
In particular, if the standard wave functions $\psi_{+}(x,t)$ and $\psi_{-}(x,t)$ are normalized we
have that
$$|\psi(x,t)|^{2}_{\bold{3}}=\frac{P_1+P_2}{2}\in [0,1]$$
where $P_1$ and $P_2$ are respectively the density probability of
$\psi_{+}(x,t)$ and $\psi_{-}(x,t)$.
\end{theorem}

\noindent \emph{Proof.}
The proof of this theorem is obtained using the following analog of the Pythagoras Theorem for bicomplex
numbers (see \cite{8}):
\be
|z_1+z_2\bold{i_2}|^2=\left|\frac{z_1-z_2\bold{i_1}}{\sqrt{2}}\right|^2+\left|\frac{z_1+z_2\bold{i_1}}{\sqrt{2}}\right|^2\mbox{ }\forall z_1+z_2\bold{i_2}\in\mathbb{T},
\label{eq:wr3.2}
\ee
to the bicomplex wave function $\psi(x,t)$. $\Box$\\

\noindent We are now ready to summarize our results with this following corollary:
\begin{corollary}
Let $\psi$ be a bicomplex wave function given by $\psi(x,t)= \\
\me^{\alpha(x,t)+\beta(x,t)\bold{i_1}+\gamma(x,t)\bold{i_2}}=
\psi_{+}(x,t)\, \bold{e_1}+\psi_{-}(x,t)\, \bold{e_2}$. Then, \be
|\psi|^{2}=|\psi|^{2}_{\bold{1}} = |\psi|^{2}_{\bold{2}}=|\psi|^{2}_{\bold{3}}
                                 = \sqrt{\psi\psi^{\dag_1}\psi^{\dag_2}\psi^{\dag_3}}
                                 = \frac{|\psi_+|^{2}+|\psi_-|^{2}}{2}
                                 = \me^{2\alpha},
\ee where $|\psi|^{2}$ gives the standard Born's formula and is
invariant under all the discrete symmetries (\ref{eq:groupe}) of
the bicomplex Schr\"odinger equation.
\end{corollary}

\smallskip\smallskip
The fact that hyperbolic angle of the exponential is zero in
$\psi$, i.e. we have to consider $\delta(x,t)=0$ in Corollary 1,
do not means that hyperbolic part of the wave function do not play
any role. Indeed, the wave function can be explicitly rewritten as
\[
\ba{rcl}
\psi(x,t)&=&\me^{\alpha(x,t)+\beta(x,t)\bold{i_1}+\gamma(x,t)\bold{i_2}}
\\*[2ex]
&=&\me^{\alpha}\left(\cos\beta \cos \gamma+ \bold{i_1}\sin \beta
\cos \gamma + \bold{i_2}\cos \beta \sin \gamma + \bold{j}\sin
\beta \sin \gamma \right). \ea
\]
Hence, the wave function considered in Corollary 1 is really a
bicomplex function.

\section{Conclusion}
In this paper we have introduced the bicomplex numbers and some
bicomplex conjugates and moduli associated with these numbers.
Then we have study the bicomplex Schr\"odinger equation where we
have found the bicomplex continuity equations. Moreover we have
shown that, under some discrete symmetries of the system of four
equations of the bicomplex  Schr\"odinger equation, the bicomplex
continuity equations can be transformed into real continuity
equations associated with the currents $J(\psi_{\pm})$. These two
real currents are in fact associated with the bicomplex
Schr\"odinger equation written in terms of the idempotent basis.
Finally, we have shown that it is possible to obtain some specific
generalization of the Born's formula for a class of wave functions
with a null hyperbolic angle. This class of wave functions are
completely invariant under all the discrete symmetries founded.

\subsection*{Acknowledgments}
The research of D.R. was supported by CRSNG of Canada. The research of S.T.
was supported by a Postdoctoral Fellowship from FQRNT du Qu\'ebec. The main part of the work here was written while S.T. was visiting the Faculty of Nuclear
Sciences and Physical Engineering of the Czech Technical University in Prague. He thanks Professor Pelantov\'a for her hospitality.

\end{document}